\documentclass{article}
\usepackage[utf8]{inputenc}
\usepackage[english]{babel}
\usepackage{graphicx}
\usepackage{subcaption}
\usepackage{amssymb}
\usepackage{dirtytalk}
\usepackage{listings}
\usepackage{csquotes}
\usepackage{amsmath}
\usepackage{amsthm}
\usepackage{algpseudocode} 
\usepackage{algorithm} 
\usepackage{xcolor}
\usepackage{dirtytalk}
\usepackage{authblk}
\usepackage{multirow}

\usepackage[margin=0.9in]{geometry}
\usepackage{hyperref}
\usepackage{lineno}

\usepackage{comment} 

\usepackage[
backend=biber,
style=numeric,
maxnames=19,
sorting=none,
citestyle=numeric
]{biblatex}
\title{Authorship Conflicts in Academia: an International Cross-Discipline Survey}

\author[1*]{Elizaveta Savchenko}
\author[2]{Ariel Rosenfeld}
\affil[1]{Department of Mathematics, Ariel University, Ariel, Israel}
\affil[2]{Department of Information Science, Bar-Ilan University, Ramat-Gan, Israel}
\affil[*]{Corresponding author: Elizavet.Savchenk@msmail.ariel.ac.il}

\date{}

\begin{document}

\maketitle

\begin{abstract}
Collaboration among scholars has emerged as a significant characteristic of contemporary science. As a result, the number of authors listed in publications continues to rise steadily. Unfortunately, determining the authors to be included in the byline and their respective order entails multiple difficulties which often lead to conflicts. Despite the large volume of literature about conflicts in academia, it remains unclear how exactly these are distributed over the main socio-demographic properties, as well as the different types of interactions academics experience. To address this gap, we conducted an international and cross-disciplinary survey answered by 752 academics from 41 fields of research and 93 countries that statistically well-represent the overall academic workforce. Our findings are concerning and suggest that conflicts over authorship credit arise very early in one's academic career, even at the level of Master and Ph.D., and become increasingly common over time.
\end{abstract}

\section{Introduction}
Scientific collaboration has become a prominent feature in modern science \cite{lee2005impact}.  
Consequently, \textit{sole authorship} (one publication—one author), which was a common practice until the twentieth century \cite{greene2007demise}, has transformed over time to \textit{co-authorship} (one publication-multiple authors) \cite{adams2013fourth}, with an ever-increasing number of authors lists on papers \cite{wuchty2007increasing,comp_18,international_collaborations}. 

Unfortunately, determining \textit{which} authors should be listed in the byline and their respective  \textit{order} encompasses multiple challenges \cite{teddy_paper,authorship_hard_1,authorship_hard_2,authorship_hard_3}. Specifically, the inherent competition for jobs, promotions, grants, and recognition amongst researchers is often associated with self-interested behavior which may lead to \textit{conflicts} over these two authorship credit distribution questions (authorship credit conflicts, for short) \cite{conf_1,conf_2,conf_3,nhb_1}. Naturally, as the number of parties involved in collaborative research work increases, the magnitude and intensity of these conflicts increase proportionally \cite{borry2006author,collaborations_conflicts,collaborations_conflicts_2}. 
 
In order to mitigate these and similar conflicts and bring about a \say{fair and transparent} authorship credit distribution, journals have adopted formal criteria to define which contributors should be listed and, in some cases, even determine their appropriate order \cite{claim_2_1}. Several common criteria include, but not limited, to ICMJE-2009\footnote{\url{http://www.ease.org.uk/sites/default/files/ease_guidelines-june2011c.pdf}}, EASE-2011\footnote{\url{http://www.icmje.org/ethical_1author.html}}, CSE-2012\footnote{\url{http://www.councilscienceeditors.org/files/public/entire_whitepaper.pdf}}, and COPE-2008\footnote{\url{http://publicationethics.org/files/u2/All_flowcharts.pdf}}. 
These criteria generally agree that authorship credit should be limited to \say{individuals who have contributed in a meaningful and substantive way to its intellectual content} \cite{claim_2_0}. However, they usually adopt somewhat vague or ambiguous terms that can lead to different interpretations. For example, it is not clear what is the minimal threshold for contribution to consider one's contribution to be \say{meaningful} enough to be listed as an author. Moreover, it is rather simple to find cases in which the different criteria disagree  \cite{claim_2_2,claim_2_2_2}. 
In parallel, several other solutions have been proposed to allow authors to better convey the role each party played in the collaboration work and the credit that should be associated with each co-author  \cite{new_name_1,jounral_2,journals_3}. One popular solution is the \textit{CRediT} (Contributor Role Taxonomy) system which requires authors to report the roles and contributions of each author separately based on fourteen different contribution types \cite{credit}. Other common solutions to indicate varying levels of contributions include multiple first authors (i.e., co-first author) and multiple corresponding authors (i.e., co-corresponding authors) \cite{multi_author}.  
Many consider these and other similar solutions to be important steps in the right direction \cite{dir_good_cite}. 

Unfortunately, there is only a very limited understanding of the extent and characteristics of the underlying phenomena today \cite{academic_issues_5,academic_issues_3,academic_issues_6}. Specifically, existing literature investigating academic conflicts has predominantly considered individual research fields, countries or very specific types of conflicts in isolation \cite{comp_17,academic_issues_1,academic_issues_2,academic_issues_4,nhb_2}. As such, the practical relevance of the proposed solutions, and perhaps the future development of novel more advanced solutions, could be significantly improved by considering more comprehensive data that accounts for multiple fields of study, nations, and types of authorship conflicts. 
In this work, we seek to fill this gap in knowledge by reporting the results of an international cross-discipline survey targeted at estimating and characterizing authorship credit conflicts in academia. 
From a methodological standpoint, by asking researchers to anonymously report their prior experiences in authorship conflicts, we are provided with a unique opportunity to reveal data that is not directly observable otherwise. Specifically, examining and analysing published outputs of collaborative work, which is the common methodological instrument practiced by prior work in this context \cite{ucl_1,ucl_2,ucl_3,ucl_4}, can only attest to the final resolution of conflicts, and only if one had actually occurred. 
Unique to our survey is the distinction between seemingly symmetrical and obviously asymmetrical relations. Specifically, our study focuses on two standard types of academic relations: one between colleagues (i.e., a seemingly symmetrical relation) and one between an academic advisor and an advisee (i.e., an inherently asymmetrical relation). To the best of our knowledge, authorship  conflicts in the advisor-advisee relationship have yet to be examined in prior literature and thus contributes a novel perspective on the matter. 

The remainder of this article is organized as follows: Section \ref{sec:methods} presents the data gathering and statistical analysis approaches. Then, Section \ref{sec:results} outlines the results followed by Section \ref{sec:discussion} which summarizes, interprets, and discusses the results in the wider context.




\section{Methods and Materials}
\label{sec:methods}
\begin{figure}
    \centering
    \includegraphics[width=0.99\textwidth]{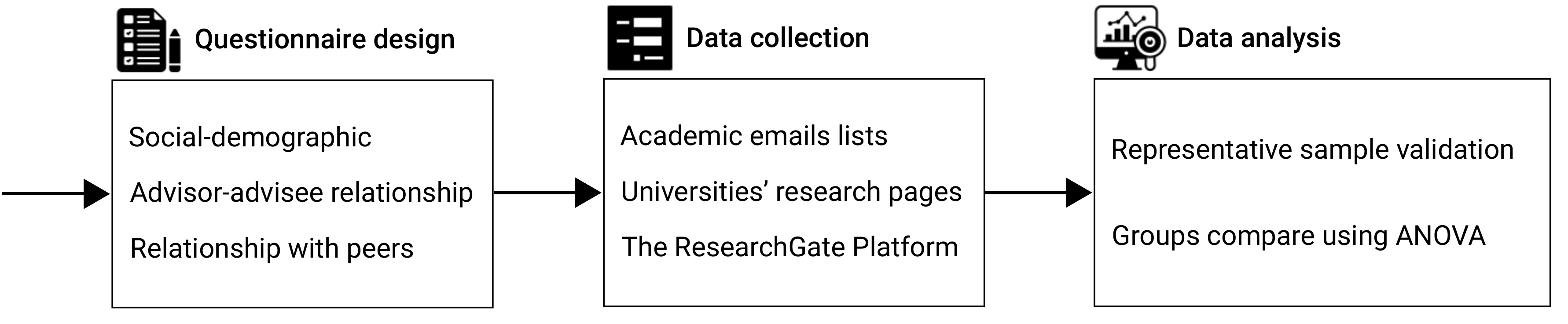}
    \caption{A schematic view of the study's methodological process.}
    \label{fig:scheme}
\end{figure}

From a methodological standpoint, we ask the scholars themselves about their behavior and experience. To gather the required data, we utilize an online survey approach. The survey was developed, distributed, and analysed as detailed below. Fig. \ref{fig:scheme} shows a schematic view of the process.

\subsection{Survey Development}
Inspired by previous studies which have developed surveys to study scholars' behaviors and preferences  \cite{comp_1,comp_2,comp_3,comp_4,comp_5}, we designed a 24-item closed-form questionnaire consisted of four parts: First, socio-demographic questions (6 items); Second, work and publication patterns-related questions (3 items); Third, questions pertaining to potential authorship credit distribution conflicts with one's Masters and PhD advisor(s) (6x2=12 items); Last, questions pertaining to potential authorship credit distribution conflicts with one's peers (3 items). The full set of items, as well as their possible closed-form answers, is provided in the Appendix.

The rationale of the survey, as presented to the respondents at the beginning of the survey, stated that \say{This survey intends to examine the extent and characteristics of disagreements in academic co-authorship.}. We chose to use the term \textit{disagreement} instead of conflict in this context in order to avoid inducing negative connotations.

\subsection{Distribution}
In order to reach as many researchers as possible we used an online questionnaire in the form of a Google form. All responses were kept confidentially and anonymously as clearly stated to the participants at the beginning of the questionnaire. The data was collected between July 2022 and November 2022. During this time, emails, as well as social media messages in the academic social media platform ResearchGate\footnote{\url{https://www.researchgate.net/}} inviting researchers to participate in the survey, were sent. The targeted researchers were chosen manually based on their affiliation country and field of research, as self-evident or self-declared in their email's signature, ResearchGate profile, or personal website in order to obtain an adequate representation of the worldwide research workforce. 
Overall, 15,362 inventions were sent and a sample of 752 researchers from 93 countries and 41 fields of research was obtained (\(4.89\%\) response rate). 
All participants had to acknowledge that they actively conduct academic research in order to avoid non-relevant samples and provided their informed consent. 
The representativeness of the obtained sample is established next.
The study was approved by the corresponding IRB.

\subsection{Representative Sample}
We first establish that the obtained sample is, indeed, representative of the studied population. For this purpose, one has to show that the sampled data satisfies two primary conditions: 1) it is large enough, and 2) the measurable parameters' distribution that is known (or approximated) for the entire population is statistically indistinguishable from these of the sample \cite{well_rep_1,well_rep_2,well_rep_3}.

First, under the assumption that all socio-demographic measures used in the questionnaire are normally distributed, we conducted a sample size power test \cite{power_test}, assuming the overall number of academics is 10.26 million\footnote{The \textit{UNESCO Science Report}\cite{unesco} has estimated 7.8 million researchers at 2013 with 21\% grows over six years, from 2007. We extrapolate these numbers by assuming that the growth rate is constant and computed a linear projection to 2022 by multiplying \(7.8\) by \(1.315\).} and a required p-value of \(0.01\). From this configuration, 667 or more samples are sufficient. Second, we analysed 22 past research papers which have used, presumably,  representative samples of the academic community. For each paper, we manually extracted the sample size and computed the mean and standard divination to be \(429 \pm 518\) with a median of \(257\). A full description of the papers used in this analysis and their sample sizes are provided in the Appendix. Taken jointly, our sample size is larger than the minimal number of required participants according to the power test, \(1.75\) times larger than the average sample size and \(2.92\) times larger than the median one. 

For the second condition, we used the age, gender, and national affiliation distributions of academics in the world as reported by the \textit{UNESCO Science Report} \cite{unesco}. We performed a Kolmogorov–Smirnov test \cite{dist_test} between the entire population's characteristics and our co-distribution of these parameters, obtaining a p-value of \(0.038 < 0.05 \). Thus, the population is statistically well represented by our sample. 

\subsection{Analytical Approach}
The comparison of subsets of the data which fulfill some condition, e.g., male vs female respondents, is conducted using the $\chi^2$ test \cite{xi_square}. When more than two groups are compared a $\chi^2$ test is conducted, followed by a post-hoc $\chi^2$ tests with Bonferroni Adjustment \cite{chi_sqaure_fix}. 
All statistical analysis was conducted using the Python programming language \cite{python}, version 3.7.5. Unless stated otherwise, significance is determined at \(p<0.05\).

Due to a large number of possible national affiliations, for our analysis, we grouped them into continents following the definition of the OECD (Organisation for Economic Co-operation and Development)\footnote{For more information please refer to \url{https://www.oecd.org/about/document/ratification-oecd-convention.htm}}.  In a similar manner, the fields of research were grouped into five \say{disciplines} - Exact, Social, Nature, Engineering, and Medicine sciences, following UNESCO's methodology \cite{unesco}.

\section{Result}
\label{sec:results}

Our analysis consists of four parts: First, we report the main socio-demographic characteristic of our sample. Then, we analyze the extent and characteristics of conflicts by considering the advisor-advisee and peer relationships, separately. Last, we consider the interaction between the two analyses -- that is, the possible link between one's conflicts as an advisee and one's subsequent conflicts with his/her peers.

\subsection{Socio-Demographics}
Our sample consists of 752 academics, \(517\) (\(68.8\)\%) of whom are male and \(233\) (\(31\)\%) are female (2 participants opted not to identify their gender). In terms of age: 3.7\% are below 25 years old, 21.7\% are 26-35 years old, 30.2\% are 36-45 years old, 24.3\% are 46-55 years old, 13.8\% are 55-65 and 6.3\% are above 65 years of age. The participants vary in their most recently obtained academic rank as follows: 0.9\% hold only a Bachelor's degree, 7.6\% hold a Master's degree, 51.2\% hold a PhD or MD, and 40.3\% have reached a rank of Professor (either associate, full or emeritus). Similarly, the participants are primarily affiliated with \(93\) different countries all across the world with the three  most prominent ones being the United States of America (USA) (\(9.3\)\%), United Kingdom  (\(8.6\)\%), and Israel (\(7.3\)\%). The participants also reported their main field of research to span over \(41\) different research subjects with the three  most prominent ones being  Mathematics (\(7.4\)\%), Economics  (\(6.8\)\%), and Informatics (\(6.4\)\%). 
Considering the participants' academic age (i.e., years passed since their first academic publication), 34.2\% report 20 or more years, 24.3\% report 10-20 years, 20.9\% report 6-9 years, 13.8\% report 3-5 years and the remaining 6.8\% report less than 3 years.

\subsection{Advisor-Advisee Conflicts}

For the following analysis, we omitted the responses of those who reported having no advisor for the relevant period. For example, individuals who pursued a direct PhD were omitted from consideration under the Master's period analysis. Overall, 26 responses (3.3\%) were omitted for Master's period analysis and 5 responses ($<$1\%) were omitted from the PhD's period analysis. Overall, in our data, the ratio between those who had a single advisor and those who had multiple advisors is 4.8 (Masters) and 2.4 (PhD), the ratio between those who had a Professor as their main advisor and those who had not is 2.3 (Masters) and 15.9 (PhD), and the ratio between those who had a male primary advisor and those who had a female one is 5.5 (Masters) and 9.9 (PhD). 
Next, we consider each of the examined characteristics of the advisor(s)-advisee relationship and the associated conflict prevalence as reported in the two relevant questions of the survey -- \say{Have you ever had a disagreement with your \textit{Masters’} advisor(s) over authorship credit distribution (i.e., who should get authorship credit or how the author byline should be ordered)} and \say{Have you ever had a disagreement with your \textit{PhD’s (or MD)} advisor(s) over authorship credit distribution (i.e., who should get authorship credit or how the author byline should be ordered)}. Table \ref{tab:advisors} summarizes the main results. 

Starting by considering all respondents together, 10.5\% have reported having at least one authorship conflict with their Masters' advisor(s), whereas 16.3\% reported the same for their PhD advisor(s). Statistically, respondents are more likely to report a conflict during their PhD with their advisor(s) compared to their Master's period, $p<0.01$. In addition, respondents who had a conflict with their Masters' advisor(s) are significantly more likely to have a conflict with their PhD advisor(s) as well, 11.4\% vs 17.2\%, $p<0.05$.

Considering the number of advisors, the results are mixed. First, having more than a single \textit{Masters} advisor is associated with a significant \textit{increase} in conflict prevalence (19.6\% vs 8.7\%) whereas having more than a single \textit{PhD} advisor is associated with a significant \textit{decrease} in conflict prevalence (17.6\% vs 13.8\%), both at $p<0.05$. In addition, having a single PhD advisor is associated with a significantly higher conflict prevalence compared to having a single Masters advisor, $p<0.01$.  

In four out of the five examined disciplines (Exact, Social, Nature and Medicine) the prevalence of conflicts with one's PhD advisor(s) is higher compared to that with Masters advisor(s). However, the difference is only statistically significant for the Exact sciences for which the relevant respondents reported  17.1\% conflict prevalence with their PhD advisor(s) compared to 11\% conflict prevalence with their Masters advisor(s), $p<0.05$. For the Engineering discipline, the opposite result is encountered with more conflicts encountered during the Masters period, yet the difference is not statistically significant. 

Turning to the issue of gender, we see that both male and female respondents are significantly more likely to report conflicts with their PhD advisor(s) compared to their Masters advisor(s). Specifically, 11.3\% and 15.8\% of all male respondents report a conflict with their Masters and PhD advisor(s), respectively, $p<0.01$. Similarly, 8.2\% and 18.4\% of all female respondents report a conflict with their Masters and PhD advisor(s), respectively, $p<0.05$. In addition, female respondents were found to have a  conflict significantly more often with their PhD advisor(s) compared to their male counterparts, $p<0.05$. Considering the primary advisor's gender, we see that having a male PhD primary advisor is associated with more conflicts than having a male Masters primary advisor, $p<0.05$. Otherwise, the primary advisor's gender is not found to significantly associate with conflict prevalence. Regarding the issue of gender concordance (i.e., the primary advisor and the respondent's gender alignment), we see that when the genders do not align, conflicts are more prevalent during the PhD period compared to the Masters period, $p<0.05$.   

Considering the primary advisor's academic title, we see that having a Professor as the primary PhD advisor is associated with a significantly higher conflict prevalence than having a Professor as the primary Masters advisor, $p<0.05$. Specifically, 16.4\% of the respondents who had a Professor as their primary PhD advisor have reported a conflict compared to 10.2\% of those who had a Professor as their primary Masters advisor.
Otherwise,  the primary advisor's academic title is not found to significantly associate with conflict prevalence.  

As for the age difference between the primary advisor and the advisee, we see that a 20-40 years difference is associated with a statistically significant increase in conflict prevalence from the Masters to the PhD period. Specifically, 19.9\% of the relevant respondents have reported a conflict with thier PhD advisor(s) compared to 9.7\% who reported a conflict with their Masters advisor(s). Otherwise,  the age difference is not found to significantly associate with conflict prevalence. 

Considering the number of published papers during the relevant period, we encounter only a single statistically significant difference. Specifically, 59.2\% of the respondents who published two papers during their Masters reported a conflict with their advisor(s) compared to only 17.3\% of the respondents who published two papers during their PhD significantly more the relevant training period, $p<0.05$. Otherwise,  the number of published papers is not found to significantly associate with conflict prevalence.  


For all six examined continents, the prevalence of conflicts with one's PhD advisor(s) is higher compared to that with Masters advisor(s). However, the difference is only statistically significant for Europe (15.4\% vs 9.3\%, $p<0.05$), whereas the difference in North America (14\% vs 5.8\%), South America (35.9\% vs 10.3\$), Africa (29.3\% vs 19.5\%) and Asia (12.6\% vs 11.3\%) is not statistically significant.

\begin{table}[!ht]
\centering
\begin{tabular}{|p{0.25\textwidth}|p{0.2\textwidth}|p{0.15\textwidth}|p{0.15\textwidth}|p{0.1\textwidth}|}
\hline 	\textbf{Characteristic} & &\textbf{Masters} & 	\textbf{PhD} & 	\textbf{p value}  \\ \hline

\multirow{3}{*}{Number of Advisors} & Single (562, 476) & 8.7\% (49) & 17.6\% (75) & \(0.003\) \\ 
&Multiple (114, 200) & 19.6\% (22) & 13.8\% (28) & \(0.298\) \\
&\textbf{p value} & \(0.030\) & \(0.019\) & - \\ \hline

\multirow{6}{*}{Advisee's Discipline} & Exact (246) & 11.0\% (27) & 17.1\% (42) & \(0.037\) \\ 
& Social (184) & 8.2\% (15) & 17.4\% (32) & \(0.052\) \\ 
& Nature (88) & 8.0\% (7) & 19.3\% (17) & \(0.188\) \\ 
&  Engineering  (108) & 16.7\% (18) & 9.3\% (10) & \(0.083\) \\ 
& Medicine (50) & 6.0\% (3) & 22.0\% (11) & \(0.418\) \\ 
&\textbf{p value} & \(0.071\) & \(0.126\) & - \\ \hline

\multirow{3}{*}{Advisee's Gender}&Male (469) & 11.3\% (49) & 15.8\% (74) & \(0.004\) \\ 
&Female (207) & 8.2\% (17) & 18.4\% (38) & \(0.026\) \\
&\textbf{p value} & \(0.069\) & \(0.040\) & - \\ \hline 

\multirow{3}{*}{Primary Advisor's Gender}&Male (580, 617) & 11.5\% (67) & 17.0\% (105) & \(0.013\) \\
&Female (96, 59) & 10.3\% (10) & 11.9\% (7) & \(0.528\) \\
&\textbf{p value} & \(0.403\) & \(0.094\) & - \\ \hline 
\multirow{3}{*}{Gender Concordance}&Yes (205, 476) & 11.0\% (23) & 15.8\% (75) & \(0.062\) \\ 
&No (471, 200) & 8.8\% (41) & 18.5\% (37) & \(0.040\) \\
&\textbf{p value} & \(0.138\) & \(0.063\) & - \\ \hline 

\multirow{3}{*}{Primary Advisor's Title}&Dr. (471, 35) & 10.7\% (50) & 20.0 \% (7) & \(0.317\) \\ 
&Prof. (205, 641)& 10.2\% (21) & 16.4\% (105) & \(0.046\) \\
&\textbf{p value} & \(0.309\) & \(0.066\) & - \\ \hline 

\multirow{7}{*}{Age difference}&Younger (2, 4) & 0\% (0) & 0\% (0) & - \\ 
&$<5$ years (7, 10) & 14.3\% (1) & 10.0\% (1) & - \\
&$5-10$ years (52,  34) & 7.7\% (4) & 14.7\% (5) & \(0.419\) \\
&$10-20$ years (306, 176) & 11.1\% (34) & 12.5\% (22) & \(0.703\) \\
&$20-40$ years (247, 412) & 9.7\% (24) & 19.9\% (82) & \(0.038\) \\
&$>40$ years (62, 40) & 11.3\% (7) & 5\% (2) & \(0.832\) \\
&\textbf{p value} & \(0.183\) & \(0.067\) & - \\ \hline

\multirow{7}{*}{Papers Published}&$0$ (300, 18) & 10.3\% (31) & 5.6\% (1) & - \\
&$1$ (294, 59) & 46.0\% (135) & 18.6\% (11) & \(0.059\) \\
&$2$ (65, 156) & 59.2\% (38) & 17.3\% (27) & \(0.038\) \\
&$3$ (17, 234) & 29.4\% (5) & 16.2\% (38) & \(0.216\) \\
&$4$ (0, 107) & 0\% (0) & 15.0\% (16) & - \\
&$5$ (0, 102) & 0\% (0) & 18.6\% (19) & - \\
&\textbf{p value} & \(0.012\) & \(0.087\) & - \\ \hline 

\multirow{6}{*}{Continent} & North America (86) & 5.8\% (5) & 14.0\% (12) & \(0.063\) \\ 
& South America (39) & 10.3\% (4) & 35.9\% (14) & \(0.057\) \\ 
& Europe (311) & 9.3\% (29) & 15.4\% (48) & \(0.040\) \\ 
& Africa (41) & 19.5\% (8) & 29.3\% (12) & \(0.093\) \\ 
&  Asia  (151) & 11.3\% (17) & 12.6\% (19) & \(0.214\) \\ 
& Oceania (49) & 14.6\% (7) & 14.6\% (7) & \(1.0\) \\ 
&\textbf{p value} & \(0.071\) & \(0.055\) & - \\ \hline

\multirow{3}{*}{Conflict during Masters}&Yes (70) & - & 17.2\% (12) & - \\
&No (606) & -  & 11.4\% (69) & -  \\ \ &\textbf{p value} & - & \(0.031\) & - \\ \hline 
Overall && 10.5\% (75) & 16.3\% (115) & \(0.008\) \\ \hline 
\end{tabular}
\caption{Characteristics of the advisor(s)-advisee relationship and the associated conflict prevalence ($N$ is given in parentheses).}
\label{tab:advisors}
\end{table}

\subsection{Peer Conflicts}

Here, we consider the respondents' characteristics and the associated conflict prevalence as reported in the three relevant questions: 1) \say{Have you ever had a disagreement with a peer over authorship credit distribution (i.e., who should get authorship credit or how the author byline should be ordered)?} (we denote this question as \textit{Conflict Faced}); 2)  \say{Have you ever had to demand more authorship credit on a paper than that was initially assigned to you (i.e., get authorship credit or improve your placement in the author byline)?} (denoted as \textit{Raised Demands}); and 3) \say{Have you ever faced a peer who raised demands to get more authorship credit than you believed s/he is entitled to (i.e., get authorship credit or improve his/her placement in the author byline)?} (denoted as \textit{Others Demand}). Table \ref{tab:peers} summarizes the main results.

Starting by considering all respondents together, we see that about one-half of all respondents have reported facing a conflict over authorship credit distribution with their peers (i.e., 48.9\% reported \say{Yes} to \say{Conflict Faced}), demanded to get more authorship credit themselves (i.e., 53.1\% reported \say{Yes} to \say{Raised Demands}) and encountered peers who demanded to get more credit than that they are entitled to (i.e., 52.9\% reported \say{Yes} to \say{Others Demand}).

Starting with the issue of gender, Male respondents reported significantly higher conflict rates in terms of Raised demands (56.1\% vs 50.2\%) and Others Demand (48.1\% vs 39.6\%) but not in Conflict Faced (50.5\% vs 54.6\%). 

Considering respondents' age, we see that older respondents tend to report higher conflict rates. Considering Conflict Faced, the conflict prevalence ranges from 26.8\% for the 25-35 age group, to 74.4\% for the above 65 age group with the prevalence rate monotonically increasing by age group, $p<0.01$. Similarly, for the Raised Demands, the prevalence ranges from 38.4\% for the 26-35 age group to 67.4\% for the above 65 age group with the prevalence monotonically increasing by  age group, $p<0.05$. Albeit not statistically significant, a similar pattern is observed for the Others Demand, with the prevalence generally increasing from 28.3\% for the 26-35 age group to 60\% for the 56-65 age group. The slight decrease in conflict prevalence observed for the above 65 age group (\say{only} 53.5\%) is presumably partially attributed with the lack of statistical significance. Similarly, considering the respondents' academic age, a similar pattern is observed with older respondents generally reporting higher conflict prevalence. Considering Conflict Faced, the conflict prevalence ranges from 23.9\% for the 3-5 age group to 72.3\% for the above 20 age group with the prevalence monotonically increasing by age group, $p<0.05$. Likewise, considering Raised Demands, the conflict prevalence ranges from 39.1\% for the 3-5 age group to 66.5\% for the above 20 age group with the prevalence monotonically increasing by age group, $p<0.05$. Again, albeit not statistically significant, a similar pattern is observed for the Others Demand with the prevalence generally increasing from 20.7\% for the 3-5 age group to 56.6\% for the above 20 age group, $p=0.06$. 

Similar to the age-based differences discussed above, the respondents' academic title is strongly associated with conflict rates as well. Specifically, Professors report higher conflict prevalence rates compared to Doctors (Conflict Faced: 71.1\% vs 39.1\%; Raised Demands: 66.3\% vs 46.2\%; and Others demand 58.4\% vs 36.9\%), all at $p<0.01$. 


Considering the respondents' workload, as represented by the number of concurrent projects pursued by the respondents, we see that a non-significant general trend where more concurrent projects are slightly associated with higher conflict prevalence. However, this trend is not statistically significant, possibly due to the extremely similar results observed for the cases of three, four and five+ concurrent projects. 

Focusing on the respondents' publication patterns, we see that the number of co-authored papers is  associated with conflict prevalence across the three examined questions, $p<0.05$. Specifically, the single co-authored paper group reported 22.9\% (Conflict Faced), 33.3\% (Raised Demands) and 12.5\% (Others Demand) compared to 67.8\%, 64.4\% and 61.9\% reported by the 10+ co-authored group, respectively. This result is naturally aligned with those presented above considering one's age, academic age and title. When considering the number of solo papers, no statistically significant differences are found. 

    
    

\begin{table}[!ht]
\centering
\begin{tabular}{|p{0.2\textwidth}|p{0.2\textwidth}|p{0.15\textwidth}|p{0.15\textwidth}|p{0.15\textwidth}|}
\hline 	\textbf{Characteristic} & &\textbf{Conflict Faced} & 	\textbf{Raised demands} & 	\textbf{Others demand} \\ \hline

\multirow{3}{*}{Gender}&Male (469) & 50.5\% (237) & 56.1\% (263) & 48.1\% (226)  \\ 
&Female (207) & 54.6\% (113) & 50.2\% (104) & 39.6\% (82) \\
&\textbf{p value} & 0.208 & 0.058 & 0.041  \\ \hline 

\multirow{7}{*}{Age}&$<25$ (4) & 0\% (0) & 0\% (0) & 0\% (0)  \\ 
&$26-35$ (138) & 26.8\% (37) & 38.4\% (53) & 28.3\% (39) \\
&$36-45$ (216) & 44.4\% (96) & 49.5\% (107) & 41.7\% (90) \\
&$46-55$ (175) & 64.0\% (112) & 64.6\% (113) & 54.9\% (96)  \\
&$56-65$ (100) & 73.0\% (73) & 65.0\% (65) & 60.0\% (60) \\
&$>65$ (43) & 74.4\% (32) & 67.4\% (29) & 53.5\% (23) \\
&\textbf{p value} & 0.009 & 0.045 & 0.060 \\ \hline 

\multirow{7}{*}{Academic Age}&$1-2$ (17)& 5.9\% (1) & 23.5\% (4) & 29.4\% (5)  \\
&$3-5$ (92) & 23.9\% (22) & 39.1\% (36) & 20.7\% (19)  \\
&$6-9$ (146) & 40.4\% (59) & 47.3\% (69) & 33.6\% (49)\\
&$10-20$ (179) & 52.0\% (93) & 54.2\% (97) & 54.7\% (98) \\
&$>20$ (242) & 72.3\% (175) & 66.5\% (161) & 56.6\% (137) \\
&\textbf{p value} & 0.014 & 0.030 & 0.062 \\ \hline 

    \multirow{4}{*}{Title}&Below Dr. (19) & 0\% (0) & 21.0\% (4) & 15.8\% (3) \\ 
&Dr. (366) & 39.1\% (143) & 46.2\% (169) & 36.9\% (135)  \\
&Prof. (291) & 71.1\% (207) & 66.3\% (193) & 58.4\% (170) \\
&\textbf{p value} & 0.001 & 0.004 & 0.004 \\ \hline

\multirow{6}{*}{\#Projects}&$1$ (78) & 23.1\% (18) & 26.9\% (21) & 15.4\% (12) \\ 
&$2$ (203) & 40.9\% (83) & 49.8\% (201) & 31.0\% (63) \\
&$3$ (185) & 61.6\% (114) & 60.5\% (112) & 55.1\% (102)  \\
&$4$ (86) & 60.5\% (52) & 65.1\% (56) & 57.0\% (49)  \\
&$5+$ (124) & 66.9\% (83) & 62.1\% (77) & 66.1\% (82)  \\
&\textbf{p value} & 0.081 & 0.099 & 0.050 \\ \hline 

\multirow{6}{*}{\#Co-authored papers}&$0$ (21) & 9.5\% (2) & 19.0\% (4) & 4.8\% (1) \\ 
&$1$ (48) & 22.9\% (11) & 33.3\% (16) & 12.5\% (6) \\
&$2-5$ (127) & 35.4\% (45) & 39.4\% (50) & 25.2\% (32)  \\
&$5-9$ (123) & 40.7\% (50) & 54.5\% (67) & 39.0\% (48) \\
&$10+$ (357) & 67.8\% (242) & 64.4\% (230) & 61.9\% (221)  \\
&\textbf{p value} & 0.029 & 0.042 & 0.011  \\ \hline 

\multirow{6}{*}{\#solo papers}&$0$ (185) & 27.6\% (51) & 42.2\% (78) & 27.0\% (50) \\ 
&$1$ (80) & 36.2\% (29) & 50.0\% (40) & 35.0\% (28) \\
&$2-5$ (133) & 66.2\% (88) & 54.9\% (73) & 63.9\% (85)  \\
&$5-9$ (125) & 61.6\% (77) & 61.6\% (77) & 52.0\% (65)  \\
&$10+$ (153) & 68.6\% (105) & 64.7\% (99) & 52.3\% (80) \\
&\textbf{p value} & 0.087 & 0.059 & 0.104 \\ \hline 

\multirow{6}{*}{Continent} & North America (86) & 51.2\% (44) & 58.1\% (50) & 47.7\% (41) \\ 
& South America (39) & 66.7\% (26) & 53.8\% (21)  & 51.3\% (20)  \\ 
& Europe (311) & 53.4\% (166) & 52.4\% (163)  & 46.0\% (143)  \\ 
& Africa (41) & 43.9\% (18) & 63.4\% (26) & 46.3\% (19)  \\ 
&  Asia  (151) & 49.7\% (75) & 51.0\% (77)  & 43.7\% (66) \\ 
& Oceania (49) & 43.8\% (21) & 62.5\% (31)  & 39.6\% (19)  \\ 
&\textbf{p value} & \(0.083\) & \(0.075\) & \(0.070\) \\ \hline

Overall && 48.9\% (368) & 53.1\% (399) & 52.9\% (398) \\ \hline 
\end{tabular}
\caption{Characteristics of the respondent and the associated conflict prevalence ($N$ is given in parentheses).}
\label{tab:peers}
\end{table}

Turning to the connection between the three questions of interest, as can be seen in Table \ref{tab:conflicts2conflicts}, the three seem to be highly interwoven. Specifically, providing a positive answer to any one of the three questions is strongly associated with increased chances of providing positive answers to the remaining two questions, $p<0.05$. For example, those who reported Conflict Faced are significantly more likely to report Raised Demands (65.7\% vs 42\%) and Others Demand (60.9\% vs 29.1\%). Similarly, those who reported Raised Demands are significantly more likely to report Others Demand (57.5\% vs 31.4\%) and vice-versa (68.5\% vs 43.4\%).

\begin{table}[!ht]
\centering
\begin{tabular}{|p{0.2\textwidth}|p{0.1\textwidth}|p{0.15\textwidth}|p{0.15\textwidth}|p{0.15\textwidth}|}
\hline 	\textbf{} && \textbf{Conflict faced}&	\textbf{Raised demands} & 	\textbf{Others demand}\\ \hline

\multirow{3}{*}{\textbf{Conflict faced}} & Yes (350) & & 65.7\% (230) & 60.9\% (213)   \\ 
&No (326) & & 42.0\% (137) & 29.1\% (95)\\
&\textbf{p value} & &  \(0.038\) & \(0.025\) \\ \hline

\multirow{3}{*}{\textbf{Raised demands}} & Yes (367) & 62.7\% (230) & & 57.5\% (211)   \\ 
&No (309) & 38.8\% (120) & & 31.4\% (97) \\
&\textbf{p value} & \(0.018\) & & \(0.016\)  \\ \hline

\multirow{3}{*}{\textbf{Others demand}} & Yes (308) & 69.2\% (213) & 68.5\% (211)   & \\
&No (368) & 37.2\% (137) & 42.4\% (155)  & \\
&\textbf{p value} & \(0.033\) & \(0.026\) &  \\\hline
\end{tabular}
\caption{Conflict prevalence across the three examined questions ($N$ is given in parentheses).}
\label{tab:conflicts2conflicts}
\end{table}

\subsection{Cross Analysis}

Last, we examine the possible link between authorship credit conflicts with one's Masters and/or PhD advisor(s) and his/her subsequent conflicts with his/her peers after graduation. As can be seen in Table \ref{tab:advisors2peers}, having a conflict with one's advisor(s) (either Masters and/or PhD) is associated with an increased rate of peer conflicts, all at $p<0.05$. Specifically, those who reported a conflict with at least one of their advsiors are significantly more likely to provide a positive answer to Conflict Faced (59\% vs 49.4\%), Raised Demands (59\% vs 52.7\%) and Others Demand (51.8\% vs 43.5\%). When breaking down the  conflicts to conflicts with one's Masters and PhD advisor(s), we see that the PhD conflicts follow the same pattern, and present a statistically significant association with one's subsequent peer conflicts. Specifically, a conflict with one's PhD advisor is associated with higher Conflict Faced (62.5\% vs 49.6\%), Raised Demands (59.8\% vs 53.2\%) and Others Demand (55.4\% vs 43.6\%). However, a conflict with one's Masters advisor(s), follows the same pattern only for Others Demand (51.4\% vs 44.9\%). While the difference in Raised demands is statistically indistinguishable, for Conflict Faced we see the opposite trend, where those who reported a conflict with their Masters advisor(s) reporting \textit{less} peer conflicts (48.6\% vs 52.1\%).

\begin{table}[!ht]
\centering
\begin{tabular}{|p{0.15\textwidth}|p{0.1\textwidth}|p{0.2\textwidth}|p{0.2\textwidth}|p{0.2\textwidth}|}
\hline 	
\textbf{Past conflict} & &\textbf{Conflict Faced} & 	\textbf{Raised demands} & 	\textbf{Others demand}   \\ \hline
\multirow{3}{*}{Masters}&No (606) & 52.1\% (316) & 54.1\% (328) & 44.9\% (272)  \\ 
&Yes (70) & 48.6\% (34) & 55.7\% (39) & 51.4\% (36)\\
&\textbf{p value} & \(0.046\) & \(0.071\) & \(0.008\)  \\ \hline 
\multirow{3}{*}{PhD}&No (564) & 49.6\% (280) & 53.2\% (300) & 43.6\% (246) \\ 
&Yes (112) & 62.5\% (70) & 59.8\% (67) & 55.4\% (62) \\
&\textbf{p value} & \(0.006\) & \(0.028\) & \(0.009\)  \\ \hline 
\multirow{3}{*}{Any}&No (510)& 49.4\% (252) & 52.7\% (269) & 43.5\% (222) \\ 
&Yes (166) & 59.0\% (98) & 59.0\% (98) & 51.8\% (86)  \\
&\textbf{p value} & \(0.002\) & \(0.027\) & \(0.040\)   \\ \hline 
\end{tabular}
\caption{Characteristics of the respondent and the associated conflict prevalence ($N$ is given in parentheses).}
\label{tab:advisors2peers}
\end{table}

\section{Discussion and Conclusions}
\label{sec:discussion}

Our results combine to suggest a complex, arguably disturbing, multifaceted picture. 

First, our results show that conflicts over authorship credit distribution are often encountered very early in one's academic career. Specifically, nearly one out of four participants have reported at least one conflict with an advisor either during their Masters and/or their PhD (24.5\%). These conflicts are especially prominent during one's PhD training period, a period which is naturally longer, mostly more productive, and possibly more competitive than the Masters training one. Interestingly, those who had a conflict with their Masters advisor(s) are less likely to have a conflict with their PhD advisor(s). A reasonable explanation could pose that the advisees who encountered a conflict with their Masters advisor(s) have chosen a different advisor(s) when pursuing their PhD. Conflict prevalence with one's advisor(s) seems to be high despite some non-consistent moderators such as the number of advisors, discipline, genders, titles, productivity, and geography. 

Authorship credit distribution conflicts seem to intensify later on in one's career, this time in the form of \say{peer conflicts}. Specifically, the results suggest that roughly one-half of the participants in this study have experienced a conflict with their peers. These conflicts seem to escalate with age and experience with older, more experienced, and more productive participants reporting higher conflict rates compared to their counterparts. Conflict prevalence with one's peers seems to be high despite some non-consistent moderators such as gender, workload, and geography.

Interestingly, those who had a conflict with an advisor during their training, either with their Masters advisor(s) and/or their PhD advisor(s), are more likely to encounter conflicts later on in their academic careers with their peers. We believe that this intriguing phenomenon can be explained in two, possibly complementary, ways. First, it may be the case that both types of conflicts are a consequence of one's personality and behavioral traits. For example, some scholars may be self-centered, egotistical, or even narcissistic, factors which are naturally associated with a more conflict-prone demeanor \cite{d_c_1,d_c_2,d_c_3}. Alternatively, conflict-prone advisors may have passed on certain academic norms, beliefs, and values to their advisees, causing them to follow a conflict-prone path, similar to how \say{parental influences} shape the development and choices of children later on in their lives \cite{d_p_1,d_p_2}. For example, an advisee may have unwittingly adopted some contentious behaviors by modeling or mimicking a past advisor, behaviors which have led to the very conflict they had with that advisor. Given the great benefits and importance of mentoring in academia \cite{qi2017standing,lienard2018intellectual,rosenfeld2022should}, a more in-depth investigation into this issue seems merited.  


Taken jointly, our results strongly suggest a systematic challenge in academia rather than a small, confined phenomenon that could be attributed to \say{a few bad apples}. Specifically, the consistency of the results across countries, fields of research, and most examined socio-demographic properties suggest that the issue is of a very large scale and should be treated as such. In addition, the fact that conflict-prone advisees \say{evolve} into conflict-prone scholars raises concerns about the current academic advisement practices.  


In future work, we intend to investigate additional academic conflicts which may arise in other interactions academics have. For example, a scholar may act as a reviewer or committee member for a grant proposal or a submitted paper for consideration. In these settings, that scholar may be tempted to leverage his/her role and act in a self-interested way by providing a negative review for a competing proposal or requesting additional references to specific works s/he authored in the past. Unfolding the unique dynamics in such settings can help establish a more complete understanding of potential conflicts in academia and can be instrumental in developing new policies to mitigate such potential conflicts. Finally, as this study is based on the self-reported past experiences of the participants,  future work can also focus on tracking conflicts as they occur and evolve over time and consider them case-by-case in a qualitative manner.

\printbibliography

\section*{Appendix}
\subsection*{Questionnaire}
In this section, we provide the questionnaire's questions and possible answers:

\begin{itemize}
    \item \say{Gender}: Male, Female, Prefer not to say.
    \item \say{Age}: 18-25, 26-35, 36-45, 46-55, 56-65, 66+ 
    \item \say{Academic degree}: First, Second, Third, Professor
    \item \say{Main field of research}: Informatics, Computer Science (CS), Environmental science, Social sciences, Philosophy, Literature, Linguistics, History, Law, Politics, Economics, Management, Sociology, Psychology, Nano/Micro science, Applied physics, Mathematics, Physics, Chemistry, Mechanical engineering, Electrical and electronic engineering, Civil engineering, Architecture and building engineering, Material engineering, Process/Chemical engineering, Neuroscience, Oncology, Genome science, Biological Science, Basic biology, Anthropology, Animal life science, Pharmacy, Basic medicine, Boundary medicine, Society medicine, Clinical internal medicine, Clinical surgery, Dentistry, Nursing.
    \item \say{How many years have passed since your first academic publication (i.e., what is your academic age)}: 0, 1-2, 3-5, 6-9, 10-20, 20+.
    \item \say{Country of primary affiliation}: a list of 196 countries from \url{https://gist.github.com/kalinchernev/486393efcca01623b18d}. 
    \item \say{Based on the last 5 years, on average, how many research projects do pursue concurrently?}: 1, 2, 3, 4, 5+.
    \item \say{How many academic papers have you published thus far with one or more co-author(s) \textit{other than your advisor(s)}?}: 0, 1, 2-5, 6-10, 10+.
    \item \say{How many academic papers have you published with no co-authors (i.e., solo papers)?}: 0, 1, 2-5, 6-10, 10+.
    \item \say{How many advisors did you have during your Masters training period?}: 1, 2, More than 2, I don't have a Masters.
    \item \say{How many papers did you publish during your Masters with your advisor(s)?}: 0, 1, 2, 3+.
    \item \say{How many advisors did you have during your Ph.D. training period?}: 1, 2, 3, 4, 5.
    \item \say{How many papers did you publish during your Ph.D. studies with your advisor(s)?}: 1, 2, 3, 4, 5+.
    \item \say{How many years your primary Masters advisor is older than you?}: Younger than me, 0-5, 5-10, 10-20, 20-40, 40+.
    \item \say{What was your primary Masters advisor's academic title?}: Dr, Prof.
    \item \say{What was your primary Masters advisor's gender?}: Male, Female. 
    \item \say{Have you ever had a disagreement with your Masters' advisor(s) over authorship credit distribution (i.e., who should get authorship credit or how the author byline should be ordered) ?}: Yes, No.
    \item \say{How many years your primary Ph.D.'s  (or MD)   advisor was older than you?}: Younger than me, 0-5, 5-10, 10-20, 20-40, 40+.
    \item \say{What was your primary Ph.D. (or MD) advisor's gender?}: Male, Female.
    \item \say{Have you ever had a disagreement with your PhD's (or MD) advisor(s) over authorship credit distribution (i.e., who should get authorship credit or how the author byline should be ordered) ?}: Yes, No.
    \item \say{What was your Ph.D.'s (or MD) primary advisor's academic title?}: Yes, No.
    \item \say{Have you ever had a disagreement with a peer  over authorship credit distribution (i.e., who should get authorship credit or how the author byline should be ordered) ?}: Yes, No.
    \item \say{Have you ever had to demand more authorship credit on a paper than that was initially assigned to you (i.e., get authorship credit or improve your placement in the author byline) ?}: Yes, No.
    \item \say{Have you ever faced a peer who raised demands to get more authorship credit than you believed s/he is entitled to (i.e., get authorship credit or improve his/her placement in the author byline) ?}: Yes, No.
\end{itemize}

\subsection*{Cohort size comparison dataset}
Table \ref{table:table:appendix_compare_cohort} summarizes 22 studies about social and political practices in academia that are based on questionnaires. 

\begin{table}[!ht]
\centering
\begin{tabular}{|p{0.6\textwidth}|p{0.1\textwidth}|p{0.1\textwidth}|p{0.1\textwidth}|}
\hline 	\textbf{Title} & 	\textbf{Group invited size} & 	\textbf{Group participant size} & 	\textbf{Source} \\ \hline
 Balancing Parenthood and Academia: Work/Family Stress as Influenced by Gender and Tenure Status & 264 & 264 & \cite{comp_1} \\ \hline 
 Retaining female postgraduates in academia: the role of gender and prospective parenthood & 249 & 249 & \cite{comp_2} \\ \hline 
 Minority Under representation in Academia: Factors Impacting Careers of Surgery Residents & 3726 & 1217 & \cite{comp_3} \\ \hline 
 Ready for careers within and beyond academia? Assessing career competencies amongst junior researchers & 727 & 727 & \cite{comp_4}  \\ \hline 
 The Influence of Gender Role Conflicts \& Academic Stress Coping Ability, and Social Support on Adaptations to College Life among Male Nursing Students & 225 & 225 & \cite{comp_5} \\ \hline 
 Ethical Problems, Conflicts and Beliefs of Small Business Professionals & 1300 & 135 & \cite{comp_6} \\ \hline 
 Academic Stress Among College Students: Comparison of American and International Students. & 392 & 392 & \cite{comp_7} \\ \hline 
 Social support from work and family domains as an antecedent or moderator of work–family conflicts? & 107 & 107 & \cite{comp_8} \\ \hline 
 Measuring Acculturation Gap Conflicts Among Hispanics: Implications for Psychosocial and Academic Adjustment & 283 & 283 & \cite{comp_9} \\ \hline 
 Work to family, family to work conflicts and work family balance as predictors of job satisfaction of Malaysian academic community & 280 & 280 & \cite{comp_10} \\ \hline 
 Marital Conflicts on Academic Performance of Secondary School Students in Port Harcourt Metropolis, Rivers State & 1951 & 195 &  \cite{comp_11}\\ \hline 
 Financial Conflicts of Interest, Disclosure, and Academic Discipline & 10 & 10 & \cite{comp_12} \\ \hline 
 Torn between study and leisure: How motivational conflicts relate to students’ academic and social adaptation & 733 & 733 & \cite{comp_13} \\ \hline 
 Perceptions of Scholars in the Field of Economics on Co-Authorship Associations: Evidence from an International Survey & 580 & 580 & \cite{comp_14} \\ \hline 
 Misuse of Coauthorship in Medical Theses in Sweden & 285 & 285 & \cite{comp_15} \\ \hline 
 Measuring dissatisfaction with coauthorship: An empirical approach based on the researchers’ perception & 2344 & 2344 & \cite{comp_16} \\ \hline 
 Coauthors’ Contributions to Major Papers Published in the AjR & 272 & 196 & \cite{comp_17} \\ \hline 
 An Examination of Ethical Research Conduct by Experienced and Novice Accounting Academics & 176 & 176 & \cite{comp_18} \\ \hline 
 Honorary Coauthorship: Does It Matter? & 195 & 127 & \cite{comp_19} \\ \hline 
 Authorship conflict in Bangladesh: an exploratory study & 100 & 100 & \cite{comp_20} \\ \hline  
 The influence of early research experience in medical school on the decision to intercalate and future career in clinical academia: a questionnaire study & 117 & 66 & \cite{comp_21} \\ \hline 
 A Study of Awareness of Authorship Criteria among Academic Plastic Surgeons & 916 & 744 & \cite{comp_22} \\ \hline \hline 
Mean \(\pm\) standard deviation & \(692 \pm 908\) & \(429 \pm 518\) &  \\ \hline 
\end{tabular}
\caption{Questionnaire-based paper about academia with their cohort sizes.}
\label{table:table:appendix_compare_cohort}
\end{table}

\end{document}